\documentclass[%
 reprint,
 bibnotes,
 amsmath,amssymb,
 aps,
]{revtex4-1}

\usepackage{graphicx}
\usepackage{textcomp}
\usepackage{dcolumn}
\usepackage{bm}

\begin{document}

\preprint{APS/123-QED}

\title{Determinants of cyclization-decyclization kinetics of short DNA with sticky ends}

\author{Jiyoun Jeong}

\author{Harold D. Kim}%
 \email{Corresponding author.\\ harold.kim@physics.gatech.edu}
\affiliation{%
 School of Physics, Georgia Institute of Technology\\
  837 State Street, Atlanta, GA 30332, USA
}%

\date{\today}

\begin{abstract}
Cyclization of DNA with sticky ends is commonly used to construct DNA minicircles and to measure DNA bendability. The cyclization probability of short DNA ($<$150 bp) has a strong length dependence, but how it depends on the rotational positioning of the sticky ends around the helical axis is less clear. To shed light upon the determinants of the cyclization probability of short DNA, we measured cyclization and decyclization rates of $\sim$100-bp DNA with sticky ends over two helical periods using single-molecule Fluorescence Resonance Energy Transfer (FRET). The cyclization rate increases monotonically with length, indicating no excess twisting, while the decyclization rate oscillates with length, higher at half-integer helical turns and lower at integer helical turns. The oscillation profile is kinetically and thermodynamically consistent with a three-state cyclization model in which sticky-ended short DNA first bends into a torsionally-relaxed teardrop, and subsequently transitions to a more stable loop upon terminal base stacking. We also show that the looping probability density (the J factor) extracted from this study is in good agreement with the worm-like chain model near 100 bp. For shorter DNA, we discuss various experimental factors that prevent an accurate measurement of the J factor.
\end{abstract}

\maketitle

\section{Introduction}
DNA under physiological conditions constantly undergoes conformational changes due to thermal fluctuations. Among those changes, bending motions coupled with twist can bring distal sites into proximity \cite{tardin2017mechanics} and impact genome packaging and gene regulation \cite{brennan2016dna,becker2018bacterial}. Some of these processes involve looped DNA segments much shorter than 500 bp, a length regime where the bending energy begins to dominate the free energy of loop formation. For example, some operons in \textit{Escherichia coli}, such as \textit{lac} and \textit{gal}, are regulated by repressor proteins that form loops as small as $\sim$100 bp \cite{cournac2013dna}. Small DNA loops can also be induced by some restriction endonucleases \cite{Gemmen2006,Laurens2012,shang2014restriction} or actively extruded by chromosome packaging motor proteins \cite{Ganji2018}. In many cases, proteins stabilize small DNA loops that spontaneously arise; therefore, it is of great importance to quantify the probability of spontaneous looping events. On the other hand, the protein complexes that bridge two distal sites of short DNA segments are subjected to a significant amount of bending and torsional stress depending on the loop geometry and size \cite{zhang2006statistical,Mulligan2015}. This stress can affect the binding affinity of the protein complexes, and thereby alter the lifetime of the looped state \cite{Laurens2012,Chen2014,yan2018protein}. Recently, small DNA loops have also been used as force sensors and applicators to study bending mechanics of DNA itself or force-dependent conformational changes of other biomolecules \cite{Shroff2008,Qu2011,Fields2013,Joseph2014,Mustafa2018}. Therefore, measuring looping and unlooping dynamics of short DNA segments can give us insights into the energetics and internal forces that govern loop-associated processes and applications.

The simplest way to form DNA loops is to use DNA with two complementary single-stranded overhangs, or sticky ends, in a reaction called cyclization. In this reaction, the sticky ends of the same DNA molecule hybridize to each other to form a ``linker'' duplex. 
To a good approximation, the cyclization (looping) rate ($k_\textrm{loop}$) is thought to be the product of two quantities \cite{Crothers1992}: (i) the effective concentration of one sticky end in the proximity of the other, which is known as the J factor ($J$), and (ii) the annealing rate constant between the two sticky ends ($k_\textrm{on}$). Therefore, if $k_\textrm{on}$ is known, the J factor can be determined by measuring $k_\textrm{loop}$. The J factor can also be predicted from polymer models as a function of length, deformability, and loop geometry. Hence, the J factor has been used as a hallmark to test and refine DNA models such as the worm-like chain model. 

Nonetheless, the experimental attempts to measure the J factor of DNA shorter than one persistence length ($\sim$150 bp) have so far been controversial. Using a ligation-based assay, the Widom group first measured the J factor of short DNA molecules \cite{Cloutier2004}. This study reported an anomalously high J factor, but the anomaly was soon proven to be an artifact due to the high concentration of ligase in a study by Vologodskii \cite{Du2005}. In a more recent study, Vafabakhsh and Ha used a ligase-free fluorescence resonance energy transfer (FRET) assay to measure the J factor of short DNA molecules in the range between 50 and 200 bp \cite{Vafabakhsh2012}. The reported J factor displayed an oscillatory pattern as a function of DNA length, which indicated that the apparent cyclization kinetics depend on the torsional degree of freedom. However, the DNA length-dependent oscillatory pattern from this FRET-based cyclization study remains puzzling because it is out of phase with that from the ligation-based cyclization study \cite{Vologodskii2013,Vologodskii2013a}. 

To shed light on this unresolved issue, we investigate how DNA cyclization and decyclization rates are influenced by the torsional degree of freedom: the rotational positioning of the sticky ends around the helical axis and base stacking between the sticky ends. Using the single-molecule FRET assay, we measured both cyclization (looping) and decyclization (unlooping) rates of short DNA ($\sim100$ bp) over two helical periods with either (1) full sticky ends that allow terminal base stacking or (2) gapped sticky ends that prevent terminal base stacking. We find that the cyclization rate varies monotonically with DNA length for both sticky-end types, whereas the decyclization rate shows length-dependent oscillation only with full sticky ends, fast at half-integer number of helical turns and slow at integer number of helical turns. Based on separately measured dissociation kinetics of sticky-ended duplexes, we attribute this kinetic difference between integer and half-integer loops to both the terminal base stacking and the shearing geometry imposed by the loop. The J factors extracted from our measured cyclization and bimolecular hybridization rates are in agreement with the worm-like chain prediction down to $\sim$90 bp despite uncertainties due to sequence and experimental condition. We also explain the origin of the oscillatory J factor reported in the previous study \cite{Vafabakhsh2012} and discuss inherent uncertainties in the experimentally derived J factor that may hamper an accurate comparison to theory, especially for DNA shorter than 100 bp.

\begin{figure*}[!htbp]
\begin{center}
\includegraphics[scale = 0.85]{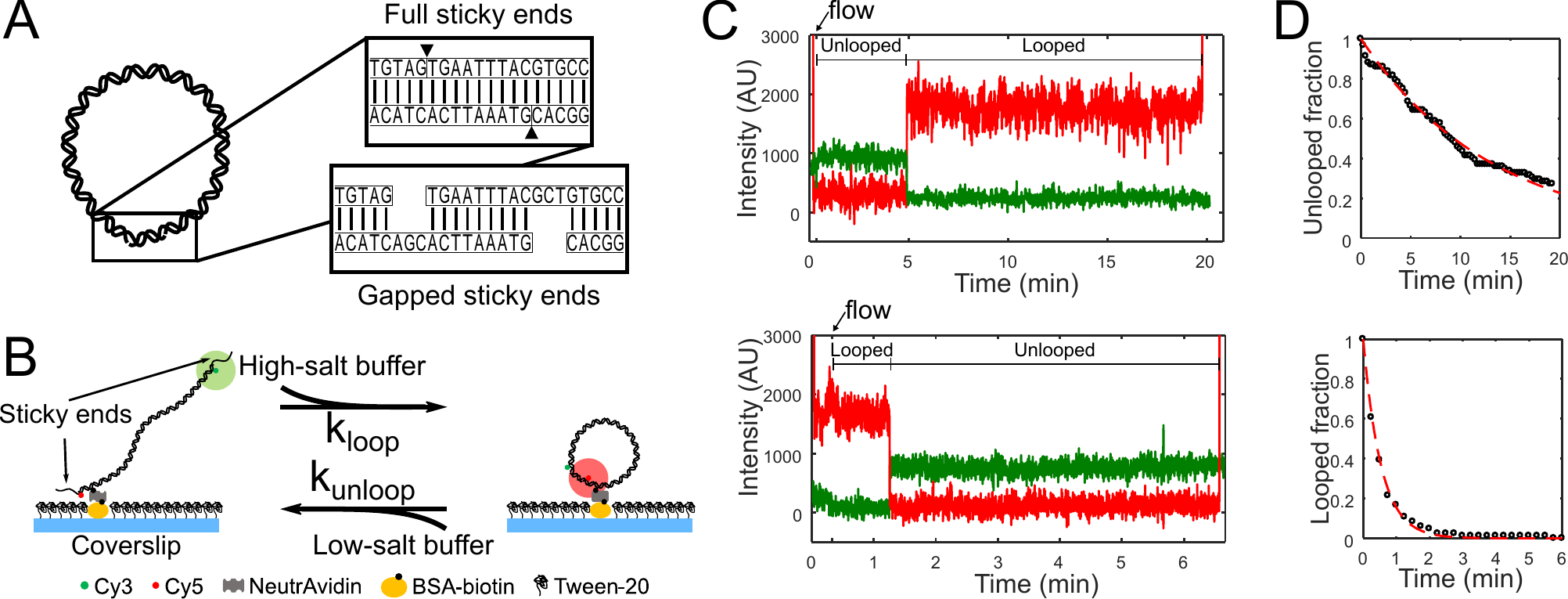}
\end{center}
\caption{\textbf{(A)} Schematic representation of a looped DNA molecule with annealed sticky ends. Close-up views show duplexed sticky ends, which we refer to as a linker duplex, without (top) and with (bottom) gaps. \textbf{(B)} Experimental setup in the FRET-based cyclization (looping) and decyclization (unlooping) assays. Fluorescently labeled DNA molecules with sticky ends are immobilized on a passivated coverslip and continuously excited by the evanescent wave of a 532-nm laser. The cation concentration of the surrounding imaging buffer is exchanged to promote either looping or unlooping of the DNA molecules. \textbf{(C)} Examples of typical fluorescence trajectories of a single DNA molecule on the surface transitioning from the unlooped state to the looped state (top) and from the looped state to the unlooped state (bottom) upon sudden salt-exchange at time = 20 s (marked by an arrow). The green and red lines represent the donor (Cy3) and acceptor (Cy5) intensities, respectively. The molecules are briefly excited by a 640-nm laser in the beginning and the end for co-localization of Cy3 and Cy5 as well as to confirm the presence of Cy5. \textbf{(D)} Examples of decay curves of the unlooped (top) and looped (bottom) fractions of molecules. The rates are extracted by fitting the data (black) with an exponential function (red).}
\label{fig1}
\end{figure*}

\section{MATERIALS AND METHODS}

\subsection{Preparation of DNA molecules with sticky ends}

We obtained two different master DNA molecules from phage lambda DNA and yeast genomic DNA by polymerase chain reactions (PCR). We performed a second set of PCR reactions on these master templates to produce DNA set 1 and set 2 whose lengths range from 96 bp to 116 bp and 108 bp to 124 bp, respectively (see Supplementary Table S1). By PCR primer design, both sets of DNA shared common 20-bp “adaptor” sequences at the ends. In each set, a DNA molecule was lengthened by inserting base pairs immediately before the two adaptor regions.

To make DNA molecules amenable to the surface-based FRET looping assay, we performed additional PCR reactions on unmodified DNA molecules with primers carrying the adaptor sequences and the necessary modifications (i.e. FRET donor, FRET acceptor, and biotin, see Supplementary Table S1 for details) \cite{Le2014jove}. Donor-labeled and acceptor-labeled double-stranded DNA molecules were made in separate PCR reactions. The donor-labeled and acceptor-labeled molecules contained the sticky-end extension at the 5$^\prime$ and 3$^\prime$-end, respectively. For gapped sticky ends, a stretch of three noncomplementary bases were inserted in the extensions (Figure \ref{fig1}(A)). The donor (Cy3) and the acceptor (Cy5) were linked to the thymine bases nearest to the 5$^\prime$ ends so that sticky-end annealing generated a high FRET signal ($\sim$0.8).

Strand exchange was performed between the two DNA molecules by incubating the mixture ($\sim$100 nM of Cy3-labeled DNA and $\sim$25 nM of Cy5-labeled DNA) at 95 $^\circ$C for 5 minutes and gradually cooling to the room temperature. As a result of strand exchange, the majority ($\sim$70 \%) of products contained all the necessary modifications as well as the 5$^\prime$ protruding sticky ends.

All of the PCR primers were commercially synthesized by Eurofins MWG Operon and Integrated DNA Technology (IDT) to at least HPLC-purity grade to minimize truncation or deletion errors. We also used Mfold \cite{Zuker2003} to ensure that each sticky end does not form unintended secondary structures.

\subsection{FRET cyclization/decyclization assay}

We adopted the previous salt-exchange FRET assay \cite{Le2014nar} except for some slight modifications in the flow-cell preparation step. We started by cleaning a microscope slide with drilled holes and a coverslip by sonication in deionized water. After sonication, the slide and the coverslip were completely dried in a vacuum chamber for about 10 to 15 minutes and etched in a plasma cleaner for additional 5 minutes. A dust-free, smooth surface was obtained at this stage. Then, we silanized the slide and the coverslip in a dichlorodimethylsilane (DDS)-hexane solution as previously described \cite{Hua2014}. After silanization, the flow-cell was assembled by joining the slide and the coverslip using double-sided tape and epoxy glue. The flow-cell was passivated and functionalized by biotinylated BSA and tween-20 before DNA molecules were injected for immobilization.

For cyclization experiments, we first incubated the molecules in an imaging buffer containing no NaCl for 10 minutes. We then started recording the time trajectories of FRET signals of the molecules and perfuse 30 uL of 1 M [NaCl] imaging buffer into the flow channel to induce looping (Figure \ref{fig1}(B)). Perfusion was controlled by a motorized syringe pump at a flow rate of 600 uL/min. The decyclization experiment was done in the same manner except that we change the salt concentration in the imaging buffer from 2 M [NaCl] to either 75 mM or 1 M [NaCl]. All imaging buffers contained the PCD-PCA oxygen scavenging system \cite{Aitken2008}. Figure \ref{fig1}(C) shows typical fluorescence intensity trajectories of Cy3 and Cy5 from these experiments. The temperature of the flow channel was maintained at 20 $^\circ$C via an objective lens temperature controller at all times. Single-molecule fluorescence data were acquired on an objective-based TIR microscope with an EMCCD camera (DU-897ECS0-\# BV, Andor) at a rate of 100 ms per frame.

\subsection{Association and dissociation rates of the linker duplex}
To measure the association rate ($k_\textrm{on}$) between the sticky ends and the lifetime ($\tau_\textrm{on}$) of the linker duplex, we prepared four different partial DNA duplexes that are sticky on one end and blunt on the other (see Supplementary Table S1). Two of them contained full sticky ends, and the other two gapped sticky ends. Each sticky end was labeled with either Cy3 or Cy5. These partial duplexes were constructed by heating a mixture of complementary oligonucleotides to 95 $^\circ$C for 5 min and gradually cooling to 4 $^\circ$C. The final concentrations of the oligonucleotides were $\sim$10 uM. The products from this reaction were purified by native polyacrylamide gel electrophoresis (12 \%, 19:1 ratio of acrylamide to bis-acrylamide in 1X TBE buffer) and extracted by ``crush and soak" followed by ethanol precipitation. The concentration of the purified product was estimated from the absorbance of the fluorescent label at its maximum absorbance wavelength. To measure $k_\textrm{on}$, one of the partial duplexes was immobilized on the surface, the other partial duplex carrying the complementary sticky end was injected into the flow cell at a known concentration, and the appearance of FRET events was monitored. To measure $\tau_\textrm{on}$, linker duplexes were formed on the surface, dissociation was induced by salt exchange, and the disappearance of FRET was monitored.

\subsection{Data analysis}

We used Matlab to extract time trajectories of FRET values from the immobilized molecules. The FRET efficiency, or signal, was calculated from the background-subtracted intensities of the donor ($I_D$) and acceptor molecules ($I_A$) using $I_A/(I_A+I_D)$. The FRET time trajectories were filtered by applying a 2-point moving average and were fed to a Hidden Markov Model estimator \cite{Bronson2009} to determine the transition points between the ideal FRET levels. The first passage time to FRET transition (low to high for looping or association and high to low for unlooping or dissociation) was collected from each FRET trajectory to build the decay curve and extract rates (see Figure \ref{fig1}(D) and Supplementary Method). 

\section{RESULTS AND DISCUSSION}

Using the single-molecule FRET assay, we measured the cyclization and decyclization kinetics of DNA near 100 bp in length. Cyclization or decyclization was triggered by a sudden increase or decrease in NaCl concentration. The FRET signals of single molecules were continuously monitored from the beginning moment of buffer exchange, and the first transition times in the FRET signals were collected to obtain mean lifetimes or rates.

\subsection{The looping rate changes monotonically, but the unlooping rate oscillates with DNA length}

\begin{figure}[t]
\begin{center}
\includegraphics[scale=0.8]{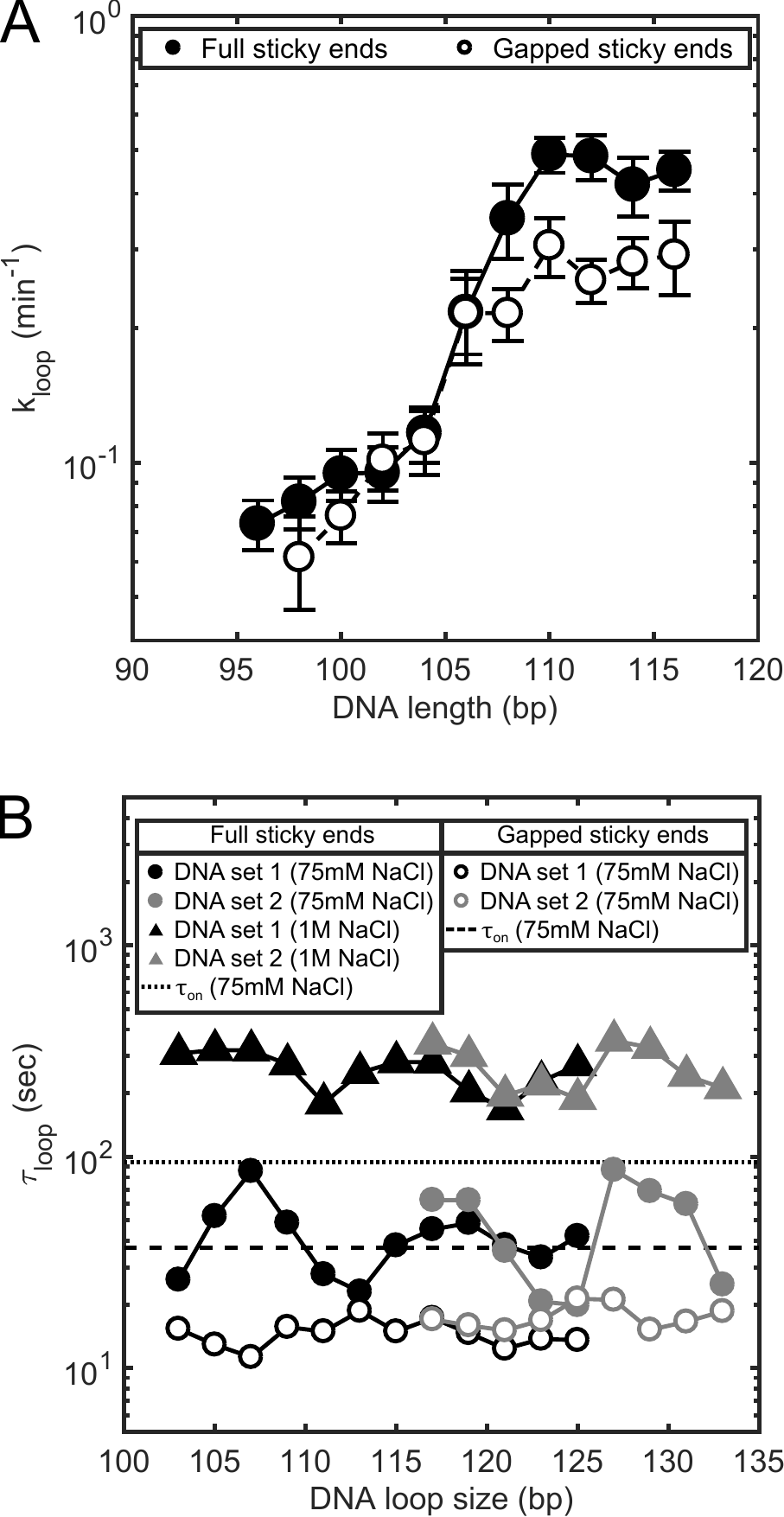}
\end{center}
\caption{\textbf{(A)} Looping rates of DNA molecules in set 1. The closed and open circles represent data measured with the full and gapped sticky pairs, respectively. Error bars represent the standard
errors of the mean. \textbf{(B)} The lifetimes of DNA molecules in DNA sets 1 and 2 in the looped state. The DNA molecules with the full sticky ends are measured in two different [NaCl] conditions: 75 mM (closed circles) and 1 M (triangles). The dotted and dashed horizontal lines represent the lifetimes of the full and gapped linker duplex, respectively. The size of error bars (not shown) is similar to the size of the data points. The DNA loop size includes the length of the annealed sticky ends (9 bp).
}
\label{fig2}
\end{figure}

In Figure \ref{fig2}(A), we present $k_\textrm{loop}$ of molecules in DNA set 1, decreasing in length from 116 bp to 94 bp in 2-bp steps. This range spans almost two helical periods of DNA. As shown, $k_\textrm{loop}$ of DNA set 1 monotonically decreases as DNA becomes shorter, indicative of the increasing energy cost of looping. The difference in $k_\textrm{loop}$ over the range of $\sim$20 bp is nearly 10-fold. The most noteworthy feature of this plot is the lack of helical-phase dependent oscillation, which shows that loop capture does not require continuity of the helical phase at the boundary. We also measured $k_\textrm{loop}$ of DNA set 1 with gapped sticky ends that prevent stacking between opposing terminal bases \cite{Kashida2017}. As expected, DNA molecules with gapped sticky ends exhibit a similar monotonic dependence of $k_\textrm{loop}$ on length. These molecules cyclize at a slightly slower rate due to the slower annealing rate of gapped sticky ends. These results indicate that stacking between the terminal bases of opposing sticky ends is not necessary for loop capture, and that the transition state must be torsionally relaxed. This is in contrast to the ligation-based assay which requires alignment of phosphate backbones, thus the oscillating cyclization rate with DNA length \cite{Shore1983,Peters2010}.  

Next, we present the dependence of decyclization kinetics on DNA length and sticky-end type. In Figure \ref{fig2}(B), we plot the lifetime of the looped state, $\tau_\textrm{loop}$, which is the inverse of the decyclization rate. In contrast to cyclization kinetics, decyclization kinetics of DNA with full sticky ends exhibit a clear length-dependent oscillatory pattern (indicated by solid symbols). The oscillation is seen with two unrelated DNA sequences (black and gray symbols) and in two different salt conditions (circle and triangular symbols). In both salt conditions, the period of oscillation is similar to one helical period of DNA (10.5 bp). At 1 M [NaCl], local maxima is identified at $\sim$105 and $\sim$115 for DNA set 1, and $\sim$127 for DNA set 2. These values are closer to integer multiples of the helical period (105, 115.5, 126 bp) than half-integer multiples. At 75 mM [NaCl], the locations of maxima (and minima) shift towards slightly larger values, which we speculate is due to curvature-dependent unwinding of a double helix \cite{Marko1994, olsen2012geometry}. The helical period ($h$) of a short DNA ring is predicted to be longer than the unstressed value ($h_0$) due to the twist-bend coupling term ($B$) according to \cite{haijun1998bending}
\begin{align}
h\approx h_0 \left( 1+\frac{1}{2}\kappa^2 \left(\frac{B}{CL}\right)^2 \right) \textrm{,}
\end{align}
where $\kappa$, $C$, and $L$ are the curvature, the torsional stiffness and the contour length of DNA, respectively. We speculate that weaker electrostatic screening at lower salt increases $B$, thus $h$ increases. Both $h_0$ and $C$ of DNA do not depend on salt \cite{taylor1990application,kriegel2017probing}. In addition to the oscillation phase, salt influences $\tau_\textrm{loop}$ and its oscillation amplitude. Loops are about 10-fold longer-lived at 1 M [NaCl] (triangles) than at 75 mM [NaCl] (circles), and the oscillation amplitude is markedly larger at 75 mM [NaCl] than at 1 M [NaCl]. 

\begin{figure}[!htbp]
\begin{center}
\includegraphics[scale=0.75]{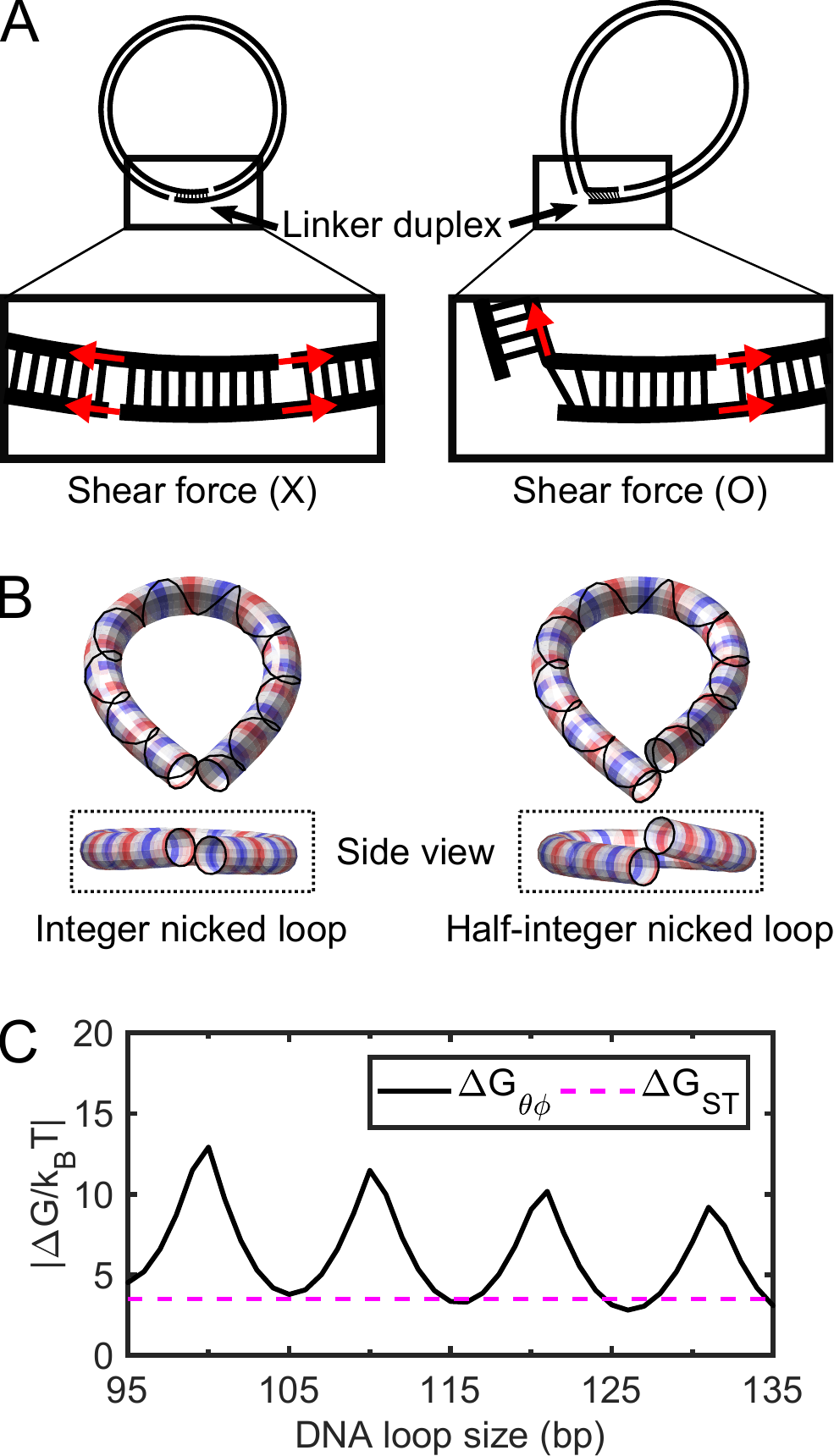}
\end{center}
\caption{\textbf{(A)} Schematic of how nick closing (terminal base stacking) can alter the stress geometry of the linker duplex. A fully stacked linker duplex (left) does not experience a shear force and therefore is more stable. In comparison, an unstacked linear duplex (right) experiences a shear force and therefore is less stable. \textbf{(B)} Minimum-energy shapes of the coarse-grained twistable worm-like chain with a single nick (left: 105 bp and right: 100 bp). Here, we consider both ends (the first and last 10 bp) of the coarse-grained chain to be cylinders with a radius equal to 1 nm whose volumes are excluded from each other during the energy minimization procedure \cite{Zhang2003}. The stand without a nick is shown as a solid line around the tubular shapes. The alternating red and blue colors indicate one helical turn (e.g. the spacing between neighboring reds (or blue) is about one helical turn). \textbf{(C)} Comparison of free energy costs. $\Delta G_{\theta\phi}$ is the free energy cost to axially and torsionally align the ends of the helix at the tip of a small teardrop loop (solid line), and $\Delta G_{\textrm{ST}}$ is the average base pair stacking energy of all 16 dinucleotides taken from Ref. \cite{Krueger2006} (dashed line). The extrapolation method in the same reference is applied to extrapolate the stacking energy for 20 $^\circ$C and [NaCl] = 0.1 M.}
\label{fig3}
\end{figure}

\subsection{The role of base stacking in the stability of DNA loop}

On the other hand, DNA loops captured with gapped sticky ends do not show length-dependent oscillation in $\tau_\textrm{loop}$ (open circles in Figure \ref{fig2}(B)). Moreover, $\tau_\textrm{loop}$ with gapped sticky ends was found to be similar in magnitude to the local minima of $\tau_\textrm{loop}$ with full sticky ends. Since the difference between full sticky ends and gapped sticky ends is the ability of base-stacking \cite{lane1997thermodynamic, Kashida2017}, we reasoned that the oscillation seen with full sticky ends arises primarily from the stacking-unstacking equilibrium at the nicks in the loop; integer loops (loops with integer number of helical turns) are longer-lived than half-integer loops because of more stable base stacking. The salt-dependent changes of $\tau_\textrm{loop}$ and the oscillation amplitude are also consistent with stabilization of base stacking at the nicks \cite{yakovchuk2006base}.

We note that base stacking at the nicks can not only provide additional stability to the linker duplex, but also dramatically alter the stress geometry of the linker duplex (Figure \ref{fig3}(A)). If one or both nicks are open due to unstacking, the linker duplex would be subject to a shear stress, which accelerates melting of short DNA duplexes \cite{Gennes2001,Whitley2016,Whitley2018}. On the other hand, if both nicks are closed as a result of stacking, the linker duplex does not experience the shear stress. To test this idea, we measured the lifetime ($\tau_\textrm{on}$) of an unstressed linker duplex produced from bimolecular association of full or gapped sticky ends. The measured lifetimes are plotted in Figure \ref{fig2}(B) as dotted are dashed lines, respectively. The amplitude of oscillation in $\tau_\textrm{loop}$ (filled circles) appears to be significantly larger than the contrast between the solid and dashed lines. This comparison reveals that stacking-dependent change in $\tau_\textrm{on}$ is not sufficient to account for the fold-change in $\tau_\textrm{loop}$ between integer and half-integer loops; therefore, the difference in the shearing geometry must be taken into account as well.    

We then ask a question as to what prevents half-integer loops from stacking at the nicks. To gain insight, we compute the minimum energy conformations of integer and half-integer loops with a single open nick \cite{Zhang2003}. In this calculation, we modeled the core of DNA as a one-dimensional twisted worm-like chain and applied the end-to-end constraint to a helical strand that winds around it. As shown in Figure \ref{fig3}(B), integer loops adopt a planar teardrop shape, whereas, half-integer loops are non-planar. Therefore, nick closing which requires axial and torsional alignment at the apex of the teardrop would be energetically more challenging to half-integer loops. We can estimate the free energy cost ($\Delta G_{\theta\phi}$) for the teardrop loop to achieve axial and torsional alignment at the nick from the J factors according to
\begin{align}
\Delta G_{\theta\phi} = - k_B T \log (J_{\theta\phi}/J)
\textrm{,}
\end{align}
where $J_{\theta\phi}$ and $J$ are the semianalytically derived J factors with and without the helical alignment, respectively \cite{Shimada1984,Douarche2005}. As predicted, $\Delta G_{\theta\phi}$ of half-integer loops is much larger than that of integer loops (solid line, Figure \ref{fig3}(C)). $\Delta G_{\theta\phi}$ of integer loop is still much larger than the thermal energy, but is comparable to the free energy ($\Delta G_{\textrm{ST}}$) of base stacking (dashed line, Figure \ref{fig3}(C)), which we estimated from the literature \cite{Krueger2006}. In agreement with our thermodynamic argument, a recent coarse-grained simulation \cite{Harrison2015} also shows that half-integer loops adopt a non-planar teardrop loop configuration in which base pair stacking across nicks is disrupted.

\begin{figure}[t]
\begin{center}
\includegraphics[scale=0.8]{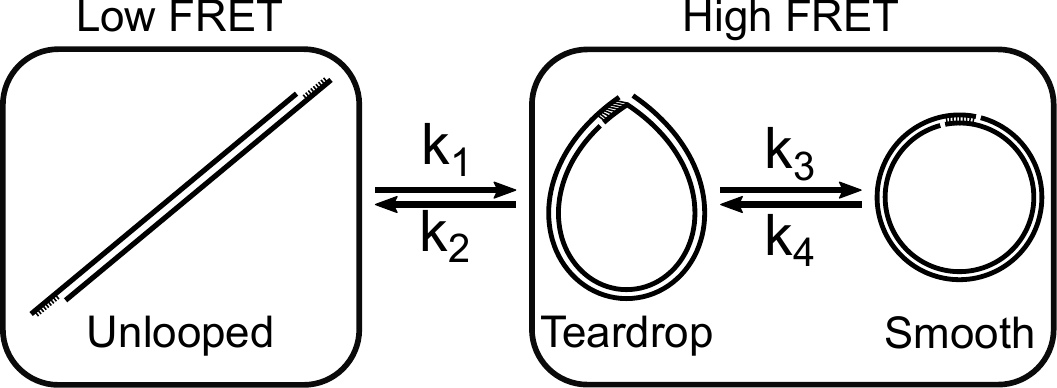}
\end{center}
\caption{Three-state DNA cyclization model. A sticky-ended short DNA molecule undergoes a transition between the low FRET (unlooped) state and the high FRET (looped) state. The transition rates between these two FRET states ($k_\textrm{1}$ and $k_\textrm{2}$) are governed by the bending energy of DNA. Two different macrostates, teardrop and smooth, can exist within the high FRET state since the looped molecule contains nicks that can spontaneously close and open. 
Transitions between the teardrop and smooth states occur at the rates of ($k_\textrm{3}$) and ($k_\textrm{4}$), respectively, and are associated with local transitions of nick closing and opening. For the transition from the teardrop state to the smooth state, integer loops need axial alignment only while half-integer loops need both axial and torsional alignment. Therefore, integer loops can transition to the smooth state more readily than half-integer loops.}
\label{fig5}
\end{figure}

Although indistinguishable by FRET, our kinetic analysis two primary macrostates in the looped state. In summary, our results validate a three-state cyclization model (Figure \ref{fig5}): (1) unlooped, (2) teardrop loop (end-juxtaposed), and (3) smooth loop (axially and torsionally aligned, and terminally stacked). A looped state with two open nicks is also possible, but is omitted from the model because it is significantly less favourable than the other two looped states (see Supplementary Results and Discussion). The oscillation-free looping rate indicates that the looped state is at first captured in a torsionally relaxed state ($k_\textrm{loop}=k_\textrm{1}$), which is likely a teardrop loop with an open nick(s). Thus, the transition rates ($k_\textrm{1},k_\textrm{2}$) between the first two states are independent of the helical phase between the two ends. In contrast, the transition rates ($k_\textrm{3}, k_\textrm{4}$) between the second and third states depend on bending, twist, and stacking energies. This equilibrium explains the difference in $\tau_\textrm{loop}$ between integer and half-integer loops. Integer loops only require in-plane bending fluctuations to close the nick, while half-integer loops require energetically demanding out-of-plane deformations to do so. Therefore, half-integer loops would be stalled in the teardrop state, and decyclize at a rate of $k_\textrm{unloop}=k_\textrm{2}$, while integer loops would be partitioned between teardrop and smooth states, and decyclize at a slower rate of $k_\textrm{unloop}=k_\textrm{2}\cdot \frac{k_\textrm{4}}{k_\textrm{3}+k_\textrm{4}}$. The single-exponential decay of the high FRET state implies that teardrop and smooth states equilibrate much faster than $k_\textrm{2}$. Based on our model, we propose the oscillation amplitude and phase in $k_\textrm{unloop}$ vs. DNA length (Figure \ref{fig2}(B)) as a useful measure to probe twist-bend coupling ($B$) and torsional stiffness ($C$) of DNA in different sequence contexts or experimental conditions. 

\subsection{Revisiting the J factor of short DNA}

The worm-like chain model is widely successful in describing the statistical mechanics of long DNA. However, whether it correctly describes the looping probability of DNA shorter than 100 bp is still debated. The comparison between measurement and model is most comprehensively shown on the plot of the J factor vs. DNA length, called the cyclization profile \cite{Tong2018}. Using the single-molecule FRET assay, Vafabakhsh and Ha \cite{Vafabakhsh2012} obtained a J factor that becomes increasingly higher than the worm-like chain prediction below 100 bp. In the same study \cite{Vafabakhsh2012}, the J factor was shown to oscillate in a length-dependent manner, and DNA with half-integer helical turns had higher J factors than DNA with integer helical turns, which is quite opposite to the results of ligation-based cyclization studies \cite{Du2005,Vologodskii2013,Vologodskii2013a}. 

As previously noted \cite{Jeong2016,Zoli2016}, the seemingly high J factor of Vafabakhsh and Ha below 100 bp is not surprising given that their J factor ($J_\textrm{VH}$) was extracted from $J_\textrm{VH}=R/k_\textrm{on}$, where $R=k_\textrm{loop}+k_\textrm{unloop}$ is an apparent relaxation rate toward the equilibrium state. Hence, $J_\textrm{VH}$ by definition is larger than $J=k_\textrm{loop}/k_\textrm{on}$ that is more closely related to the theoretical J factor. The larger discrepancy between $J_\textrm{VH}$ and theoretical J at shorter lengths is also expected because $k_\textrm{unloop}$ increases steeply with decreasing length \cite{Le2014nar}. Our new results from this study also offer a clear explanation to the out-of-phase oscillatory profile of $J_\textrm{VH}$. Although $k_\textrm{loop}$ changes monotonically with length, $k_\textrm{unloop}$ oscillates with peaks at half-integer helical turns. Therefore, $J_\textrm{VH}$, which is proportional to $k_\textrm{loop}+k_\textrm{unloop}$, would exhibit peaks at half-integer turns.

\begin{figure}[t]
\begin{center}
\includegraphics[scale=0.8]{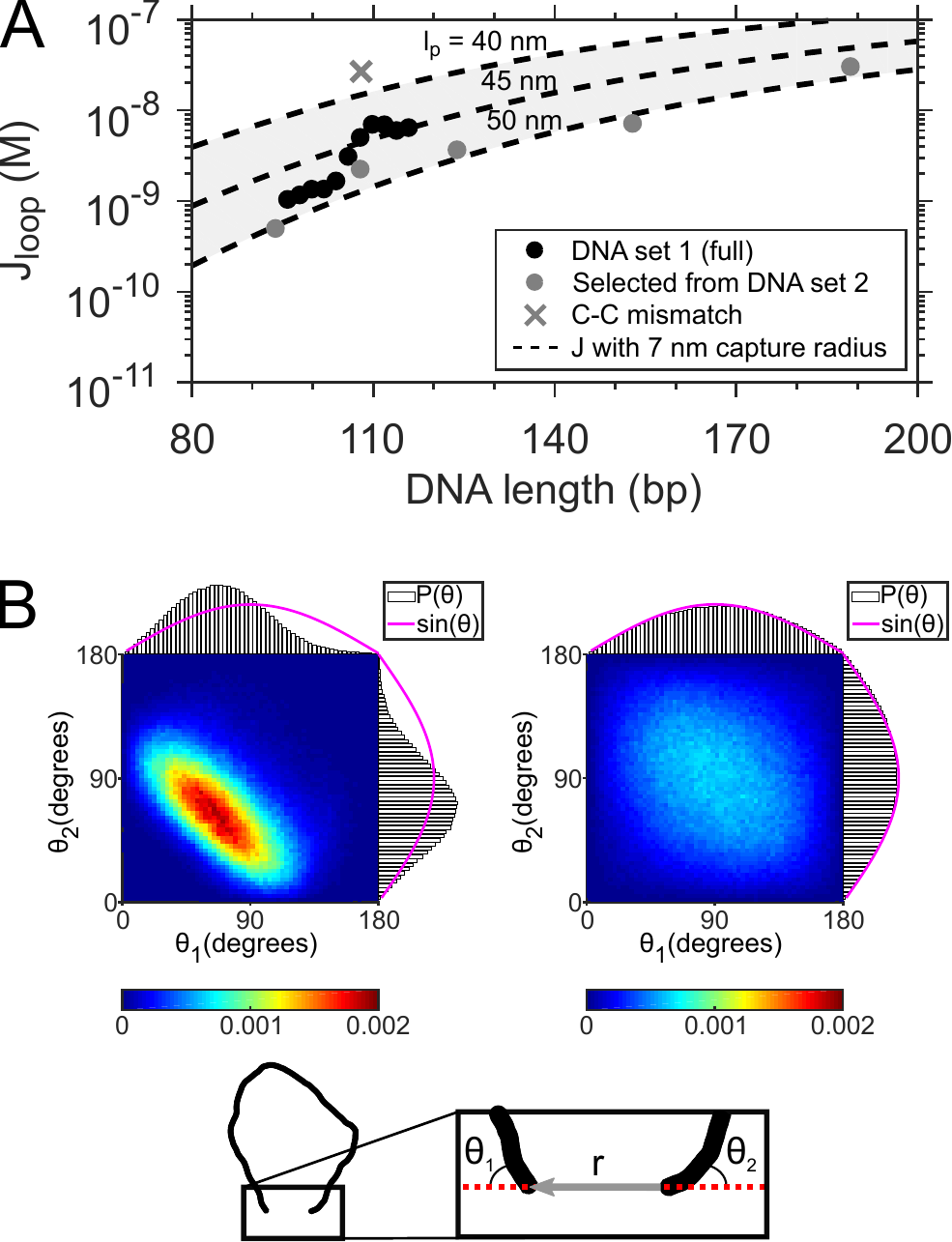}
\end{center}
\caption{
\textbf{(A)} $J_\textrm{loop}$ as a function of DNA length. The length-dependence in the extended range outside of DNA set 1 is measured with DNA molecules from set 2 (grey). The measured $J_\textrm{loop}$ is compared with the worm-like chain model prediction of the J factor (dashed lines) calculated based on Ref. \cite{Douarche2005}. In this calculation, we assumed the loop capture radius is equal to the contour length of 10-nt single-stranded DNA since loop capture is initiated by base pairing between single-stranded sticky ends. The shared area between the dashed lines represent the prediction made with a range of persistence lengths from 40 to 50 nm. 
\textbf{(B)} Joint probability distributions ($P(\theta_1,\theta_2)$) of coarse-grained DNA chains. The schematic at the bottom shows a DNA chain constrained with a short end-to-end distance, $\vert \textbf{r} \vert$. $\theta_1$ and $\theta_2$ are the angles between the chain ends and the end-to-end vector. The left and right density plots represent the joint distributions of $\theta_1$ and $\theta_2$ for 100-bp and 500-bp loops, respectively. The projected probability distributions of $\theta_1$ and $\theta_2$ are individually plotted along the x- and y-axis of each density plot, respectively. The magenta line represents the unconstrained $P(\theta_1,\theta_2)$, which is equal to the sine function.
}
\label{fig4}
\end{figure}

Following the correct expression of the J factor ($J=k_\textrm{loop}/k_\textrm{on}$), we extracted the J factors of DNA set 1 with full sticky ends using $k_\textrm{loop}$ from Figure \ref{fig2}(A) and $k_\textrm{on}$ measured from the bimolecular association of the full sticky ends. The results are shown in Figure \ref{fig4}(A) (black dots). In the same figure, we also plot the theoretical J factor of a worm-like chain \cite{Douarche2005} using a range of persistence lengths from 40 to 50 nm (dashed lines). This theoretical J factor should be taken as a lower limit as it approximates the fluctuations about the minimum-energy loop only up to the quadratic terms \cite{giovan2015dna}. As shown in this plot, the J factors of DNA set 1 correspond to persistence lengths between 44 and 49 nm. This 5-nm variability is still within the accepted range of experimentally determined values \cite{brunet2015dependence}. We also measured $J_\textrm{loop}$ from molecules in DNA set 2 over a wider length range (grey circles, Figure \ref{fig4}(A)). $J_\textrm{loop}$ from these molecules shows an overall good agreement with $J$ of 50-nm persistence length. The difference in $J_\textrm{loop}$ between the two DNA sets is consistent, but not remarkable considering that the J factor can vary with sequence by a few orders of magnitude in the similar length range \cite{Rosanio2015}. In comparison, we show that a single base pair mismatch in the center of a 108-bp DNA can increase $J_\textrm{loop}$ by almost 10-fold (cross(x) in Figure \ref{fig4}(A)). Hence, apart from sequence-dependent irregularities, our J factor measurements are consistent with the canonical worm-like chain model in the length range tested.   

\subsection{Limitations of the J factor below 100 bp}
Although understanding energetics of DNA looping at even shorter lengths ($<$100 bp) is of growing interest, we argue that the J factor is neither a theoretically relevant nor an experimentally accessible quantity in this regime. As DNA becomes shorter, end segments are more flexible than internal segments \cite{wu2015flexibility}, and end base pairs are more prone to fraying \cite{cong2015revisiting,Harrison2015,joffroy2017rolling}. Moreover, discreteness of base pairs and sequence-dependent effects cannot be sufficiently averaged out over several helical turns. Therefore, the J factor which describes the average behavior of a continuous, homogeneous polymer is no longer relevant in this length scale. 

For short DNA, the experimental $J_\textrm{loop}$ also becomes a bad proxy for the theoretical J factor. The underlying assumption in $J_\textrm{loop}=k_\textrm{loop}/k_\textrm{on}$ is that the second-order annealing rate constant between the two sticky ends in cyclization is the same as that in bimolecular association. This assumption allows the use of $k_\textrm{on}$ measured from the bimolecular reaction to cancel out the annealing rate constant $f$ in $k_\textrm{loop}$ and recover the looping probability density $J$:
\begin{align}
J_\textrm{loop}=\frac{k_\textrm{loop}}{k_\textrm{on}} = \frac{f}{k_\textrm{on}}J=J\textrm{.}
\end{align}
However, if the annealing rate constant depends on the relative orientation of the sticky ends, $f$ depends strongly on DNA length. To highlight this effect, we plot the joint probability distribution of two angles ($P(\theta_1,\theta_2)$) formed between the end-to-end vector and the helical axes of the end segments (Figure \ref{fig4}(B)) using a Monte Carlo simulation of a worm-like chain \cite{Jeong2016}. These angles thus represent how much the two sticky ends would have to deviate from the helical axes for annealing. Large angles will incur some energetic penalty because dangling bases in the sticky ends can stack \cite{Freier1985,Bommarito2000}, albeit weakly. The two angles at which two separate molecules encounter would be independent and uniformly distributed, and therefore, $P(\theta_1,\theta_2)$ should be proportional to $\sin \theta_1 \sin \theta_2$ (magenta lines, Figure \ref{fig4}(B)). A similar distribution is obtained for the ends of a 500-bp DNA, much longer than the persistence length (right, Figure \ref{fig4}(B)). For the ends of short DNA (100 bp), however, the two angles are highly restrained because of the strong bending stress in the looped DNA (left, Figure \ref{fig4}(B)). Compared to the bimolecular case, small angles favorable for annealing occur more frequently while extreme angles unfavorable for annealing occur less. It is thus conceivable that at the same end-to-end distance, $f$ would be larger than $k_\textrm{on}$, leading to higher $J_\textrm{loop}$ than the theoretical J factor.

The J factor measurement of short DNA suffers from practical complications as well. As DNA becomes shorter, the fraction of molecules that loop on a laboratory time scale becomes extremely small, and detection of this trace amount in a bulk ligation assay becomes quite laborious and cumbersome \cite{Du2005}. In our single-molecule FRET assay, extremely slow events are inevitably masked by photobleaching of the fluorophores, which leads to an overestimation of $k_\textrm{loop}$. This overestimation becomes more severe as cyclization becomes slower. In our experience, slower cyclization kinetics is also fitted more poorly with a single exponential function, possibly due to an increasing inactive fraction over time (see Supplementary Method). Therefore, $J_\textrm{loop}$ of DNA shorter than 100-bp carries substantial experimental and statistical uncertainties. In our opinion, $J_\textrm{loop}$ of short DNA should be interpreted only as a comparative measure of loopability, but not as a proxy for the theoretical J factor.

\section{CONCLUSION}
The single-molecule FRET assay \cite{Vafabakhsh2012,Le2013} can detect cyclization intermediates without the need of protein-mediated ligation and is thus thought to be a more accurate method to measure the intrinsic looping probability of short DNA. However, exact boundary interactions and loop geometry of these intermediates are not known, which complicates the interpretation of the apparent cyclization rate. In this study, we measured cyclization and decyclization rates of short DNA as a function of DNA length. The cyclization rate changes monotonically with DNA length without helical-phase dependent oscillation. In contrast, the decyclization rate showed length-dependent oscillation, faster at half-integer helical turns, and slower at integer-helical turns. We further demonstrate that the oscillation results from stackable bases at the nicks, and the oscillation amplitude can be explained by shear-accelerated nick opening. We present a three-state cyclization model that is kinetically and thermodynamically consistent with our data and existing stacking free energy parameters, and propose the oscillation profile of the decyclization rate as a new measure to explore twist and bending mechanics of DNA. Lastly, the J factors extracted from cyclization rates of 90 to 120-bp DNA are in good agreement with persistence lengths in the range of 44 to 49 nm.

\bibliography{references}
\end{document}


\title{Determinants of cyclization-decyclization kinetics of short DNA with sticky ends}
\date{}
\author{%
Jiyoun Jeong and Harold D. Kim\,%
\footnote{Corresponding author. Email: harold.kim@physics.gatech.edu}\\
{\it School of Physics, Georgia Institute of Technology, 837 State Street, Atlanta, GA 30332-0430, USA}}

\maketitle

\begin{longtable}{|c|X|}
\hline

\multicolumn{2}{|c|}{\begin{tabular}[c]{@{}l@{}}\textbf{DNA set 1 (5\textrm{$^\prime$} to 3\textrm{$^\prime$})}\end{tabular}}     \\ \hline
\endfirsthead

\hline 
\multicolumn{2}{|c|}{\begin{tabular}[c]{@{}l@{}}\textbf{DNA set 2 (5\textrm{$^\prime$} to 3\textrm{$^\prime$})} -- continued from previous page\end{tabular}}     \\ \hline
\endhead

\hline \multicolumn{2}{|r|}{{Continued on next page}} \\ \hline
\endfoot

\hline
\caption{List of DNA sequences, PCR primers, and blocking oligonucleotides. All molecules in DNA sets 1 and 2 include the common adapter sequences (20 bp) at both ends, which are also present in the PCR primer pairs. PCR primers and blocking oligos are hybridized to each other to make the partial duplexes that are sticky on one end and blunt on the other: Blocking-Cy5full and Blocking-Cy5gap are hybridized with the forward primers of full and gapped sticky ends, and Blocking-Cy3 hybridizes with the backward primers of both full and gapped sticky ends, respectively. These partial duplexes are used to measure the association rate ($k_{\textrm{on}}$) between the sticky ends and the lifetime ($\tau_{\textrm{on}}$) of the linker duplex.
} \label{tab:sequences}
\endlastfoot

94 bp                  & \seqsplit{GTGCCAGCAACAGATAGCCACCGGAGCCACACCGGTGCAAACCTCAGCAAGCAGGGTGTGGAAGTAGGACATTTCCCATTCGAGCTCGTTGTAG}                       \\ \hline
96 bp                  & \seqsplit{GTGCCAGCAACAGATAGCCATCCGGAGCCACACCGGTGCAAACCTCAGCAAGCAGGGTGTGGAAGTAGGACATTTTCCCATTCGAGCTCGTTGTAG}                     \\ \hline
98 bp                  & \seqsplit{GTGCCAGCAACAGATAGCCATTCCGGAGCCACACCGGTGCAAACCTCAGCAAGCAGGGTGTGGAAGTAGGACATTTTCCCCATTCGAGCTCGTTGTAG}                   \\ \hline
100 bp                 & \seqsplit{GTGCCAGCAACAGATAGCCACTTCCGGAGCCACACCGGTGCAAACCTCAGCAAGCAGGGTGTGGAAGTAGGACATTTTCACCCATTCGAGCTCGTTGTAG}                 \\ \hline
102 bp                 & \seqsplit{GTGCCAGCAACAGATAGCCAACTTCCGGAGCCACACCGGTGCAAACCTCAGCAAGCAGGGTGTGGAAGTAGGACATTTTCATCCCATTCGAGCTCGTTGTAG}               \\ \hline
104 bp                 & \seqsplit{GTGCCAGCAACAGATAGCCAAACTTCCGGAGCCACACCGGTGCAAACCTCAGCAAGCAGGGTGTGGAAGTAGGACATTTTCATGCCCATTCGAGCTCGTTGTAG}             \\ \hline
106 bp                 & \seqsplit{GTGCCAGCAACAGATAGCCATAACTTCCGGAGCCACACCGGTGCAAACCTCAGCAAGCAGGGTGTGGAAGTAGGACATTTTCATGTCCCATTCGAGCTCGTTGTAG}           \\ \hline
108 bp                 & \seqsplit{GTGCCAGCAACAGATAGCCATTAACTTCCGGAGCCACACCGGTGCAAACCTCAGCAAGCAGGGTGTGGAAGTAGGACATTTTCATGTCCCCATTCGAGCTCGTTGTAG}         \\ \hline
110 bp                 & \seqsplit{GTGCCAGCAACAGATAGCCAGTTAACTTCCGGAGCCACACCGGTGCAAACCTCAGCAAGCAGGGTGTGGAAGTAGGACATTTTCATGTCACCCATTCGAGCTCGTTGTAG}       \\ \hline
112 bp                 & \seqsplit{GTGCCAGCAACAGATAGCCACGTTAACTTCCGGAGCCACACCGGTGCAAACCTCAGCAAGCAGGGTGTGGAAGTAGGACATTTTCATGTCAGCCCATTCGAGCTCGTTGTAG}     \\ \hline
114 bp                 & \seqsplit{GTGCCAGCAACAGATAGCCAGCGTTAACTTCCGGAGCCACACCGGTGCAAACCTCAGCAAGCAGGGTGTGGAAGTAGGACATTTTCATGTCAGGCCCATTCGAGCTCGTTGTAG}   \\ \hline
116 bp                 & \seqsplit{GTGCCAGCAACAGATAGCCAAGCGTTAACTTCCGGAGCCACACCGGTGCAAACCTCAGCAAGCAGGGTGTGGAAGTAGGACATTTTCATGTCAGGCCCCATTCGAGCTCGTTGTAG} \\ \hline

\multicolumn{2}{|c|}{\begin{tabular}[c]{@{}l@{}}\textbf{DNA set 2 (5\textrm{$^\prime$} to 3\textrm{$^\prime$})}\end{tabular}}     \\ \hline

94 bp                  & \seqsplit{GTGCCAGCAACAGATAGCCACATGGCAACGAGGTCGCACACGCCCCACACCCAGACCTCCCTGCGAGCGGGCATCCCATTCGAGCTCGTTGTTG}                       \\ \hline
108 bp                  & \seqsplit{GTGCCAGCAACAGATAGCCACGATCGCCATGGCAACGAGGTCGCACACGCCCCACACCCAGACCTCCCTGCGAGCGGGCATGGGTACACCCATTCGAGCTCGTTGTAG}                     \\ \hline
110 bp                  & \seqsplit{GTGCCAGCAACAGATAGCCAGCGATCGCCATGGCAACGAGGTCGCACACGCCCCACACCCAGACCTCCCTGCGAGCGGGCATGGGTACAACCCATTCGAGCTCGTTGTAG}                   \\ \hline
112 bp                 & \seqsplit{GTGCCAGCAACAGATAGCCACGCGATCGCCATGGCAACGAGGTCGCACACGCCCCACACCCAGACCTCCCTGCGAGCGGGCATGGGTACAATCCCATTCGAGCTCGTTGTAG}                 \\ \hline
114 bp                 & \seqsplit{GTGCCAGCAACAGATAGCCAGCGCGATCGCCATGGCAACGAGGTCGCACACGCCCCACACCCAGACCTCCCTGCGAGCGGGCATGGGTACAATGCCCATTCGAGCTCGTTGTAG}               \\ \hline
116 bp                 & \seqsplit{GTGCCAGCAACAGATAGCCACGCGCGATCGCCATGGCAACGAGGTCGCACACGCCCCACACCCAGACCTCCCTGCGAGCGGGCATGGGTACAATGTCCCATTCGAGCTCGTTGTAG}             \\ \hline
118 bp                 & \seqsplit{GTGCCAGCAACAGATAGCCAACGCGCGATCGCCATGGCAACGAGGTCGCACACGCCCCACACCCAGACCTCCCTGCGAGCGGGCATGGGTACAATGTCCCCATTCGAGCTCGTTGTAG}           \\ \hline
120 bp                 & \seqsplit{GTGCCAGCAACAGATAGCCACACGCGCGATCGCCATGGCAACGAGGTCGCACACGCCCCACACCCAGACCTCCCTGCGAGCGGGCATGGGTACAATGTCCCCCATTCGAGCTCGTTGTAG}         \\ \hline
122 bp                 & \seqsplit{GTGCCAGCAACAGATAGCCACCACGCGCGATCGCCATGGCAACGAGGTCGCACACGCCCCACACCCAGACCTCCCTGCGAGCGGGCATGGGTACAATGTCCCCCCATTCGAGCTCGTTGTAG}       \\ \hline
124 bp                 & \seqsplit{GTGCCAGCAACAGATAGCCACCCACGCGCGATCGCCATGGCAACGAGGTCGCACACGCCCCACACCCAGACCTCCCTGCGAGCGGGCATGGGTACAATGTCCCCCCCATTCGAGCTCGTTGTAG}     \\ \hline
153 bp                 & \seqsplit{GTGCCAGCAACAGATAGCCAGGGGAAAGACCACACCCACGCGCGATCGCCATGGCAACGAGGTCGCACACGCCCCACACCCAGACCTCCCTGCGAGCGGGCATGGGTACAATGTCCCCGTTGCCACAGAGACCCCCATTCGAGCTCGTTGTAG}   \\ \hline
189 bp                 & \seqsplit{GTGCCAGCAACAGATAGCCACAGCAACGGGCAACCGTTTGGGGAAAGACCACACCCACGCGCGATCGCCATGGCAACGAGGTCGCACACGCCCCACACCCAGACCTCCCTGCGAGCGGGCATGGGTACAATGTCCCCGTTGCCACAGAGACCACTTCGTAGCACAGCGCCCCATTCGAGCTCGTTGTAG} \\ \hline

\multicolumn{2}{|c|}{\begin{tabular}[c]{@{}c@{}}\bf Full sticky ends PCR primer pairs (5\textrm{$^\prime$} to 3\textrm{$^\prime$})\end{tabular}} \\ \hline
Forward                   & TGAATTTACG{[}Cy5dT{]}GCCAGCAACAGA{[}BiotindT{]}AGC
\\ \hline
Backward                  & GTAAATTCAC{[}Cy3dT{]}ACAACGAGCTCGAATGGG
\\ \hline
\multicolumn{2}{|c|}{\bf Gapped sticky ends PCR primer pairs (5\textrm{$^\prime$} to 3\textrm{$^\prime$})}                                           \\ \hline
Forward                   & TGAATTTACGCTG{[}Cy5dT{]}GCCAGCAACAGA{[}BiotindT{]}AGCCA                             \\ \hline
Backward                  & {[}Cy3{]}GTAAATTCACGACTACAACGAGCTCGAATGGG                 \\ \hline

\multicolumn{2}{|c|}{\begin{tabular}[c]{@{}c@{}}\bf Blocking oligos for making partial duplexes (5\textrm{$^\prime$} to 3\textrm{$^\prime$})\end{tabular}} \\ \hline
Blocking-Cy5full                   & GCTATCTGTTGCTGGCAC
\\ \hline
Blocking-Cy5gap                  & TGGCTATCTGTTGCTGGCAC
\\ \hline
Blocking-Cy3                   & CCCATTCGAGCTCGTTGTAG                             \\ \hline

\end{longtable}

\section{Supplementary Method: Extracting rates from the decay curves}
The looping and unlooping rates were extracted from fitting an exponential function to the looping and unlooping decay curves. We observed that the unlooping decay curve reached to a zero within our typical recording time ($\sim$ 6 min), and thus used an exponential function of the form $N(t) = \textrm{exp}(-k_\textrm{unloop}t)$ for fitting. However, the looping decay curve did not reach to a zero even after 20 minutes of observation, which was the longest recording time we have tried in this study. According to our previous study \cite{Le2013}, we noted a certain fraction of DNA molecules, $N_\infty$, did not loop even after a long time ($\sim$40 min). Therefore, we assumed that all DNA samples contain a similar fraction of inactive molecules and fitted decay curves with an equation of the form $N(t) = (1-N_\infty)\textrm{exp}(-k_\textrm{loop}t)+N_\infty$, with $N_\infty = N_\infty(\textrm{189bp})$, which was determined from 189-bp-long DNA molecule from DNA set 2 whose decay curve quickly plateaus to $N_\infty(\textrm{189bp})$. The mean lifetimes, $\tau_\textrm{unloop}$ and $\tau_\textrm{loop}$, can be found by taking the reciprocal of $k_\textrm{loop}$ and $k_\textrm{unloop}$, respectively.

\section{Supplementary Results and Discussion: Free energy difference between the singly-kinked and doubly-kinked loops}
Replacing $J_{\phi\theta}$ with the J factor estimated with an end-to-end distance equal to 3 nm (i.e. the contour length of the linker duplex) in Equation (2) in the main text, we can calculate the free energy difference between the singly and doubly-kinked loops. Taking into account $\Delta G_{\textrm{ST}}$ ($\approx\textrm{3.5 } k_BT$), we find that the doubly-kinked loop is thermodynamically less favorable than the singly-kinked loop by $\sim\textrm{2 } k_BT$, which corresponds to an equilibrium fraction of 0.14, in the range between 95 bp and 125 bp. Therefore, we focus on our discussion on transitions between the teardrop state with a single open nick and the smooth state.

\bibliography{supplementary_ref}